\documentclass{article}
\usepackage{graphicx}
\usepackage{booktabs}
\usepackage{amsmath,amssymb}
\usepackage[inline]{enumitem}
\usepackage{hyperref}
\usepackage[T1]{fontenc}
\usepackage[accepted]{mlsys2025}
\makeatletter\renewcommand{\Notice@String}{Preprint.}\makeatother
\mlsystitlerunning{\smash{Placing LLM Lookup Tables by Prefetch Deadlines}}
\begin{document}
\twocolumn[
\mlsystitle{Placing LLM Lookup Tables by Per-Layer Prefetch Deadlines}
\begin{mlsysauthorlist}
\mlsysauthor{Wen Huang}{ms}
\mlsysauthor{Jack Liu}{ind}
\end{mlsysauthorlist}
\mlsysaffiliation{ms}{Microsoft}
\mlsysaffiliation{ind}{Independent Researcher}
\mlsyscorrespondingauthor{Wen Huang}{wenhuang@microsoft.com}
\mlsyskeywords{LLM serving, Engram, memory tiering, NVMe, prefetching}
\vskip 0.3in
\begin{abstract}
DeepSeek-V4.1's 203\,GB of Engram n-gram tables sit in host DRAM today, taking 190 of the 223\,GiB a 4-GPU serving process uses.
Row indices depend only on input tokens, so reads can start with each step and overlap earlier layers.
V4.1's two Engram layers have windows that differ by 30$\times$ (0.3 vs.\ 9.3\,ms).
We show that this deadline, together with session reuse and contention on a shared device, decides which tier each table can use and which layers need cache, and we give a model that predicts the resulting slowdown from measured read times.
Agent and chat traffic reuse rows within sessions, and static hot sets transfer poorly.
We build SlackTier, a table tier for SGLang that issues all lookups at step start.
It reads the short-window layer first and splits a small GPU row cache across layers.
The path runs inside the decode CUDA graph and is bit-exact with the original gather.
Layer~14's table can go to NVMe uncached with no measurable decode cost.
With both tables on four NVMe disks, mean time per output token on agent and chat replays rises at most 1.4\% over unmodified SGLang, and host memory falls from 212 to 34\,GiB.
Both tables need no cache when read over RDMA (remote direct memory access) from another node's memory; one copy serves two instances.
On a KV-constrained agent workload, using the freed memory as host KV cache at the same memory budget raises median throughput by 36\% and cuts median p90 time to first token from 38 to 24\,s.
\end{abstract}
]
\printAffiliationsAndNotice{}

\section{Introduction}

DeepSeek-V4.1 ships with two hash tables totaling 203\,GB.
Two of its 40 layers, layer~1 and layer~14, run an Engram module~\cite{cheng2026engram,deepseekv41}.
For every token, each module hashes the last few tokens into 24 rows of a 384M-row table and mixes those rows into the residual stream.
The tables are too big for HBM, so today's serving stacks keep them in host DRAM and pull rows over PCIe every step~\cite{lmsysv41}.
On our 4$\times$H200 node, serving uses 223\,GiB of host memory, 190\,GiB for the tables alone.
On unified-memory systems such as DGX Spark, a host table takes memory from the GPU.
Developers report that such tables make models fail to load on 128\,GB nodes when pinned in host memory~\cite{sglangqwenple,vllm62238}, and we observe the same for Qwen3.8 on a DGX Spark (\S\ref{sec:disc}).
Recent models also use lookup tables to add capacity cheaply~\cite{cheng2026engram,huang2025overtok,yu2025scemb}, so this footprint will grow.

The obvious fix is to move the tables to local NVMe.
Done naively, this slows decode by 6--7\% at 128 concurrent sessions.
Caching layer~14's hot rows on the GPU reduces the overhead only from 6.4\% to 4.9\%.

\begin{figure}[t]\centering
\includegraphics[width=\linewidth]{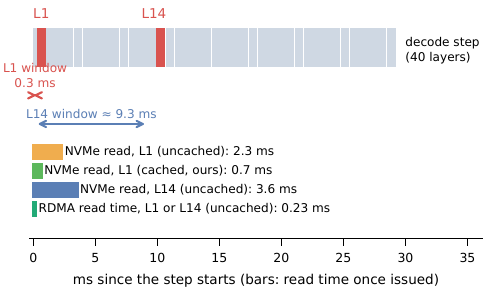}
\caption{One V4.1 decode step at 128 sessions. Both Engram lookups are issued at the start of the step. Layer~1 needs its rows after about 0.3\,ms, and layer~14 after about 9.3\,ms. An uncached NVMe read takes 2--4\,ms, so it only hurts layer~1. An uncached RDMA read from another node's memory takes about 0.2\,ms.}
\label{fig:timeline}\end{figure}

Engram's indices depend only on input tokens, so all rows for a step are known at its start, and the reads can overlap the preceding layers~\cite{ma2026cxlengram}.
The two layers get very different amounts of overlap.
The fourteen layers before layer~14 give it about 9\,ms, enough to cover a full uncached NVMe read of 2--4\,ms for a 128-session batch.
Layer~1 has 0.3\,ms, which is too short (Figure~\ref{fig:timeline}).
Placement therefore depends on each layer's read deadline as much as on which rows are hot.
Ma et al.\ state this zero-stall condition for one CXL (Compute Express Link) memory pool and note that early layers are tight~\cite{ma2026cxlengram}; deployments still put both tables on one tier~\cite{aetherixengram,plirdma}, and a DGX Spark port caches rows of both layers on NVMe and sees larger layer-1 stalls~\cite{miaaipr7}.
We measure both windows on the released model and turn the condition into a per-layer choice of tier and cache.

We present \textbf{SlackTier}, a serving tier that places each lookup table, and divides a small GPU cache, by the deadline of the layer that reads it. The idea behind it is a simple rule: \emph{a table can live on any tier whose read time fits the window of the layer that reads it, and the cache should go to the layers whose windows are too short.} We use Ma et al.'s per-layer zero-stall condition across heterogeneous tiers, together with session reuse and contention on a shared device, to decide where each table goes and which layers get the cache; hot-row caching alone ignores the window.

The rule holds up in practice. For V4.1, layer~14's table goes to NVMe with no cache and no measurable decode cost, while layer~1 needs a fast tier or a share of the cache: moving a 1\,GB cache from layer~14 to an even split cuts the overhead from 4.9\% to 1.3\%. A small cache is enough because rows are reused within sessions: a 1M-row LRU hits 74\% of agent lookups, while static hot sets built offline hit 29--51\%. When both tables share a disk, layer~14's reads also delay layer~1's, and a model that puts both layers on one clock captures this. Given each run's measured read times, with windows from the probe and one constant fitted on early runs, it predicts the decode slowdown of 174 later runs, on NVMe and RDMA and from 8 to 256 sessions, to 0.22\,ms RMSE.

Against unmodified SGLang, SlackTier with four NVMe disks costs at most 1.4\% per output token on our agent and chat replays and cuts the serving process's host memory from 212 to 34\,GiB. Reading the tables over RDMA from another node needs no cache at all, and one copy can serve two instances. On a KV-bound agent workload, reusing the freed memory as host KV cache raises median throughput by 36\%. The windows follow the same rule on three model families and three GPU platforms, so measuring a new table's window and read time under the target load tells where it can live: Qwen3.8's per-layer n-gram table sits at layer~1 like V4.1's first table, and behaves the same way.

Our contributions are:
\begin{itemize}
\item We measure per-layer prefetch windows on three model families and platforms. We measure Engram access patterns on the released V4.1 tables under agent and chat traffic (\S\ref{sec:char}, \S\ref{sec:eval}).
\item A placement rule for token-indexed tables that decides tier and cache per layer, with a simple exposure model for lookup layers that share a device (\S\ref{sec:model}).
\item SlackTier is an SGLang tier over NVMe or RDMA. It runs inside the decode CUDA graph and is bit-exact with the original gather (\S\ref{sec:design}).
\item We evaluate DeepSeek-V4.1 and Qwen3.8 against unmodified SGLang, including remote and shared tables (\S\ref{sec:eval}).
\end{itemize}

\section{Background}

\subsection{Engram in DeepSeek-V4.1}
DeepSeek-V4.1-Flash~\cite{deepseekv41} is a 40-layer mixture-of-experts (MoE) model with two Engram~\cite{cheng2026engram} modules, in layers 1 and 14.
Engram is one of several designs that give transformers large, sparsely read parameter tables, along with product-key memories~\cite{lample2019pkm,berges2024memory} and n-gram and over-tokenized embeddings~\cite{roy2022ngrammer,huang2025overtok,yu2025scemb}.
These tables add capacity at almost no compute cost, so their memory footprint determines how they are served.
For each token $t$, an Engram layer maps the 2-, 3- and 4-gram ending at $t$ through a compressed vocabulary (99{,}092 entries).
For each order, it hashes the n-gram with 8 independent heads.
Each head uses a multiplicative XOR hash modulo a distinct prime above $1.6\times10^{7}$.
This gives \textbf{24 row indices per token per layer} into a table of 384M rows.
A row stores 256 FP8 values plus eight E8M0 block scales (264\,B).
Engram concatenates the 24 dequantized rows, projects them with a dense $6144\to25600$ matrix, and adds the result through a per-token sigmoid gate to the residual stream.
The two tables total 203\,GB, larger than the rest of the checkpoint's per-GPU footprint under four-way tensor parallelism (TP4).

Two properties matter for serving.
(i) \emph{Indices depend only on input tokens}, not on hidden states.
All 48 rows a token needs are therefore known before the forward pass starts.
(ii) \emph{Rows are read-only and sparse}: a decode step with 128 sequences touches about 3{,}000 rows per layer, 0.8\,MB out of 100\,GB.

\subsection{How Engram is served today}
SGLang offers a host-table mode that keeps each table in pinned host DRAM.
It stores one shared copy or one shard per TP rank.
A Triton kernel gathers rows over PCIe through unified virtual addressing.
vLLM provides similar host and shared-memory placements.
Community proposals prefetch layer~14 on a side stream during layers 2--13, some with a GPU hot-row cache for layer~14.
Other proposals add disk-backed tables via memory-mapped safetensors, Mooncake-backed remote tables, and a unified table abstraction.
\citet{ma2026cxlengram} place the tables in a CXL memory pool.

In our TP4 deployment with four-way expert parallelism (EP4) on 4$\times$H200, host-table mode uses \textbf{223\,GiB} of resident memory for the serving process, including 190\,GiB for the two tables (\S\ref{sec:eval}).
Gathering rows from DRAM takes 20--160\,$\mu$s per decode step, under 1\% of a 17--30\,ms step, so the problem is memory, not time.
This paper asks how much of that memory can be freed, and at what cost.

\section{Characterizing Engram Accesses}
\label{sec:char}

We reproduce V4.1's Engram hash exactly on CPU and check row counts and indices against the checkpoint and SGLang's gather.
We replay three workloads: OpenHands~\cite{wang2024openhands} agent trajectories (SWE-rebench; holdout repositories never seen when fitting statistics), SWE-agent trajectories (one per issue), and WildChat~\cite{zhao2024wildchat} multi-turn chat.
We interleave sessions (64--256 concurrent) to mimic a serving node.
Prefill tokens warm the cache, and we report hit rates over \emph{decode} lookups, which are on the latency-critical path.

\paragraph{Reuse within sessions.}
A per-layer row LRU of 1M rows (264\,MB, 0.26\% of a table) hits 74\% of agent decode lookups.
At 5M rows it hits 81\% (OpenHands) to 85\% (SWE-agent).
At 5\% of the table, it hits 86--89\% (supplementary appendix).
Warming the cache with other repositories does not help (86.1\% vs.\ 85.7\%).
Chat reaches only 35\% at 1M rows, and 42--47\% from 5M rows up to 5\% of the table.
The remaining misses are first-seen n-grams: identifiers, paths and tool output specific to the current task, or long-tail 3/4-grams in chat.
Growing the cache from 1\% to 5\% of the table adds only 3--8 points.

\paragraph{Static hot sets.}
At 1M and 5M rows, a hot set from the same distribution (our fit split) covers only 51--59\% of agent lookups.
A public hot list built from 1.1B tokens of assistant traffic~\cite{botlabhot} covers 29--39\% at these sizes, even though it reaches 43--59\% on its own held-out data.
Tripling the data used to pick the hot set raises coverage by only 2--6 points.
An S3-FIFO-style policy~\cite{yang2023s3fifo} and LRU are within about 2.5 points of each other (Table~\ref{tab:hit}), so the replacement policy matters far less than admitting rows dynamically.
Pinning a static head in front of an LRU helps only at large capacities (+2--3 points at 20M rows).

\paragraph{Prediction.}
We tried three ways to hide misses without caching them.
Prefetching rows of drafted next tokens~\cite{leviathan2023spec}, with a prompt-lookup drafter~\cite{saxena2023pld} as a proxy, recovers only 2--7\% of lookups, because the drafter mostly fails on the same new tokens.
Offline packing stores each frequent 4-gram context's 24 rows contiguously, costing two pages instead of 24.
It removes 22\% of page reads for agents and 6--8\% for chat, saturating at 50\,GB (agent) to 200\,GB (chat) of extra disk.
Reading only some of the 8 hash heads does not work either. Their value contributions are nearly orthogonal (cosine $\approx0$), and dropping one changes its n-gram order's contribution by $\sim$35\%.
Within-step prefetch needs no prediction, because the current tokens' indices are known when the step starts.

\paragraph{Reuse within a batch.}
Among 256 decode tokens sampled from the trace, 95--99.6\% of rows are distinct.
In an 8K-token window, 33\% (agent) to 71\% (chat) are.
Deduplication helps prefill, and the per-row cache helps decode.

\paragraph{Qwen3.8.}
We reproduce the hash of Qwen3.8-Flash-Next (architecture class \texttt{Qwen4Exp} in SGLang and vLLM) and see the same behavior in its per-layer n-gram embedding (320M rows, 2/3-grams $\times$ 8 heads, consumed at layer~1).
Agent LRU hit rates are 86--87\% and chat 54--60\%, depending on capacity.

\begin{table}[t]
\centering\footnotesize\setlength{\tabcolsep}{3pt}
\caption{Decode-lookup hit rate (\%) of one Engram layer (layer 14) under 64 interleaved sessions.}
\label{tab:hit}
\resizebox{\columnwidth}{!}{%
\begin{tabular}{lrrrr}
\toprule
 & \multicolumn{2}{c}{Agent (OpenHands)} & \multicolumn{2}{c}{Chat (WildChat)} \\
Policy & 1M rows & 5M rows & 1M rows & 5M rows \\
\midrule
LRU & 74.0 & 81.0 & 34.8 & 41.7 \\
S3-FIFO-style & 74.5 & 78.7 & 32.7 & 39.2 \\
Static (same dist.) & 51.3 & 59.5 & 17.7 & 26.4 \\
Static (public list) & 29.2 & 39.0 & 8.0 & 12.3 \\
\bottomrule
\end{tabular}}
\end{table}

\section{Prefetch Windows and Placement}
\label{sec:model}

\subsection{Exposure model}
Let both Engram layers issue their lookups when a decode step starts, with all times measured from that point on one clock.
Layer $\ell$'s \emph{prefetch window} $W_\ell$ is the compute time before it needs its rows, following \citet{ma2026cxlengram}.
Let $s$ be the CPU time to resolve the batch's indices against the cache directory.
Let $T_\ell$ be the time from $s$ until layer $\ell$'s missing rows arrive on the slowest rank.
Layer~1's reads are submitted first, at $s$, and layer~14's right after, so $T_{14}$ includes any time layer~14 waits behind layer~1 on a shared disk.
The waits are sequential. A stall at layer~1 delays everything after it and gives layer~14 more time:
\begin{equation}
\begin{aligned}
d_1 &= [\,s + T_1 - W_1\,]^+,\\
d_{14} &= [\,s + T_{14} - W_{14} - d_1\,]^+,\\
\Delta &= d_1 + d_{14} + \kappa,
\end{aligned}
\label{eq:exposure}
\end{equation}
where $[x]^+=\max(0,x)$ and $\kappa$ is a fixed per-step cost.
V4.1 has 40 layers and takes 29.7\,ms per step at 128 sessions.
$W_{14}$ spans about 13 layers, roughly 10\,ms, while $W_1$ covers only layer~0 and the embedding.

On NVMe, page reads account for most of $T_\ell$.
We fit $T_1$ across our end-to-end configurations:
$T_1 \approx 0.30\,\text{ms} + 3.05\,\mu\text{s}\cdot P_1 + 0.37\,\mu\text{s}\cdot P_{14}$ ($R^2=0.997$).
Here $P_\ell$ counts the 4\,KiB pages layer $\ell$ reads in the step.
Because layer~1 is submitted first (\S\ref{sec:design}), layer~14's pages slow it only through contention for the shared disk, at one eighth of the per-page cost.

We calibrate $W_1$ and $\kappa$ on 24 end-to-end runs (agent and chat, nine cache configurations).
The fit gives in-sample $R^2=0.96$ (RMSE 0.14\,ms) with $W_1\approx0.6$\,ms.
In every run, $T_{14}$ stays at least 6\,ms below $W_{14}$, so $d_{14}=0$ and the recurrence reduces to the layer-1 term.
Layer~14 never binds in these runs, so the fit says little about $W_{14}$.
A direct probe (\S\ref{sec:eval}) gives $W_{14}\approx9.3$\,ms, in line with the compute of the layers before it, and $W_1\approx0.3$\,ms.
The fit determines only $\kappa-W_1$, which explains its larger $W_1$.
We use the measured windows from here on.
With the probe's windows (scaled by each batch's DRAM step time) and $\kappa$ set once on 14 early runs, Eq.~\ref{eq:exposure} predicts the mean decode slowdown of 174 later runs it never saw, on NVMe and over RDMA, from 8 to 256 sessions, from their measured per-run mean read times (averaged across ranks): RMSE 0.22\,ms over slowdowns of $-$0.3 to 3.9\,ms ($R^2=0.93$). It also predicts injected delays at both layers to 0.12\,ms (\S\ref{sec:eval}).
In these NVMe runs layer~14 never binds, so the model reduces to layer~1's excess read time plus $\kappa$; the coupling term matters only when both layers are slow, which we test by injecting delays (Table~\ref{tab:both}).

\subsection{What the model implies}
\paragraph{Placement.} A table can move to a slower tier without exposed stalls if its reads fit its window without delaying the other layer's, regardless of how often its rows are used.
For V4.1, layer~14's table can stay on NVMe uncached, but layer~1's reads would stall.
\paragraph{Cache allocation.} When both tables use NVMe behind a GPU row cache of $C$ rows, $P_1$ drives exposure, with a smaller penalty from $P_{14}$.
The layers share the disk, so layer~14's pages slow layer~1's reads even though the I/O path submits layer~1 first.
The split $f$ of the budget given to layer~1 should minimize
$b_1\,P_1(fC) + b_{14}\,P_{14}((1-f)C)$,
where $P_\ell(c)$ is the measured miss-page curve of layer~$\ell$ at cache size $c$, and $b_1/b_{14}\approx 8$.
Each $P_\ell$ falls steeply for the first rows and then flattens.
The best split therefore gives layer~1 most of the budget but not all of it. Past some point, a row moved to layer~1 saves fewer pages than layer~14 loses.
The model puts the best split around $f\approx0.8$ for agents and near 1 for chat.
In practice the curve is flat. Layer-1 shares from 0.25 to 1 differ by a few tenths of a millisecond, and all are far better than a layer-14-only cache (\S\ref{sec:eval}).
An even split performs close to the best measured allocation.
\paragraph{Generality.} The same reasoning applies to any table indexed only by input tokens.
A table used at layer $k$ has a window of roughly $k$ layers.
Qwen3.8's PLE sits at layer~1, in the same position as V4.1's layer~1, and therefore needs a fast tier or a cache.
A table added to the input embedding has no window, as with LongCat's n-gram embedding.
It must stay in a fast tier or rely entirely on caching.
A table split across all layers, as with Gemma~4's per-layer embeddings, can be split by column.
Shallow layers' columns can use fast tiers, and deep layers' columns can use slower ones.

\section{Design and Implementation}
\label{sec:design}

We implement SlackTier in SGLang~\cite{zheng2024sglang} 0.5.22 in about 900 lines of C and 300 lines of Python, without changing the model definition.

\paragraph{On-disk layout.}
Each TP rank owns a contiguous quarter of each table's rows, matching SGLang's per-rank host layout.
Rows are stored as 264\,B records, 15 per 4\,KiB page.
Each rank stores each layer's rows in a region of a raw block device opened with \texttt{O\_DIRECT}.
Ranks map to devices round-robin.
With one disk, all four ranks share it; with four disks, each rank has its own.

\begin{figure*}[t]\centering
\includegraphics[width=0.82\linewidth]{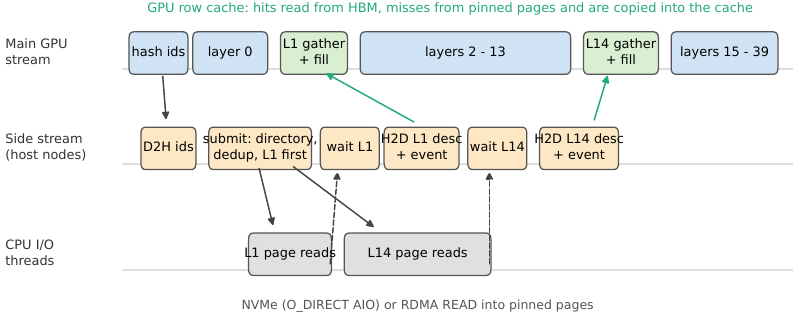}
\caption{One decode step. All host nodes, copies and events sit on a side stream inside the decode CUDA graph. Layer~1's reads are submitted before layer~14's; each gather waits only for its own layer's event.}
\label{fig:arch}\end{figure*}

\paragraph{Issuing lookups at step start.}
After the model computes hash indices, a copy to pinned host memory and a CUDA host-function node run on a side stream.
The node walks the batch's 48 indices per token.
If a row is in the GPU cache, it records the slot.
If the row was already requested in this step, it reuses that read.
Otherwise it allocates a page buffer for the row's page, unless that page is already being read, and a cache slot to fill.
With lossy skipping enabled, rows that were never seen are left as zero.
The node processes layer~1 first and hands its pages to an I/O thread \emph{before} processing layer~14, so layer~1's reads go out first.
Further host nodes wait for each layer's reads, copy its per-lookup descriptors to the GPU, and record an event.
The main stream waits on layer~1's event just before layer~1 and on layer~14's event just before layer~14.
Every operation is a host node, a copy or an event, so the whole path is captured in SGLang's decode CUDA graph and replay needs no extra launches.

\paragraph{Gather and fill.}
A Triton kernel reads each descriptor, which holds a row address in the pinned page pool or a cache slot.
The kernel dequantizes FP8 with E8M0 scales exactly as the original gather does.
On a miss, it also writes the raw 264\,B row into its newly allocated cache slot.
The cache stores packed rows in two GPU arrays of 256\,B and 8\,B per slot.
Output is bit-identical to the host-table gather.
An in-server checker that compares every row against the original kernel found no mismatch over 790K lookups, including configurations with evictions.

\paragraph{Cache directory.}
The CPU-side directory holds a row-to-slot array per layer (4\,B per row of the rank's shard).
It runs CLOCK replacement over slots and pins slots used in the current step.
One CPU pass decides hits, deduplicates pages and builds the I/O list.
This takes 80--160\,$\mu$s per step at 128 sessions and overlaps with layer~0.

\paragraph{Configuration.}
A control file sets disk count, per-layer cache sizes and each layer's source (NVMe, DRAM or RDMA) at runtime.
Our experiments use it to switch configurations within one server process.

\section{Evaluation}
\label{sec:eval}

\paragraph{Setup.}
We run DeepSeek-V4.1-Flash with SGLang 0.5.22 and TP4/EP4 on 4$\times$H200 (one node), using FP8 KV cache, decode CUDA graphs, and bounded SWA replay.
Engram tables are on local NVMe SSDs (3.5\,TB each, raw block devices).
Clients replay multi-turn sessions using OpenHands agent trajectories from held-out repositories and WildChat conversations.
Each turn sends the original history plus the new user turn and generates 128 tokens, so successive turns reuse the radix prefix cache.
$B$ denotes concurrent sessions, not a fixed decode batch.
We report ``step time'' (the mean interval between consecutive decode steps with no intervening prefill), its percentiles, and streaming TTFT.
The cache can be resized at runtime up to a reserved maximum (2M rows per layer and rank in earlier runs, 0.5M in later ones).
Unless noted, it uses 1M rows per rank ($\approx$1\,GB across the node), and a deployment would reserve only that.
The baseline uses SGLang's host-table mode with the table in DRAM, served through our gather path so every configuration shares one code path.
It is bit-identical to the stock kernel.
Step time isolates the table tier's decode cost.
Client-side results include prefill and appear in \S\ref{sec:client}.
Unmodified SGLang's processes held 212\,GiB of host memory, almost all for the two tables.
Unless noted, each configuration runs 30\,s of warm-up and 60\,s of measurement, three times in shuffled order, on one server process; the cache-size sweep, the single-layer probes and the no-graph runs use one run each.

\subsection{Memory and overhead}
Resident memory falls from \textbf{223\,GiB} with the host table to \textbf{34\,GiB} with the tables on NVMe.
Table~\ref{tab:main} shows the step-time overhead and host memory of each placement.

\begin{table*}[t]
\centering\small
\caption{Decode step time at $B{=}128$ relative to the host-DRAM table, step times from one server process. Mean of three shuffled runs (run-to-run spread $\le$0.16\,ms); p99 step time in parentheses. ``1 disk'' is one NVMe shared by the four ranks; ``4 disks'' is one per rank with four I/O threads each. Cache: rows per rank, split across the two layers unless noted ($\approx$1\,GB total for 1M).}
\label{tab:main}
\begin{tabular}{lrrrr}
\toprule
Configuration & Host memory & Cache & Agent & Chat \\
\midrule
Host-DRAM table & 223\,GiB & --- & 29.41\,ms (31.1) & 33.71\,ms (35.1) \\
Layer~1 DRAM, layer~14 on 1 disk & $\sim$129\,GiB & --- & $-$0.0\% (31.1) & $-$0.1\% (35.1) \\
Both on 1 disk & 34\,GiB & --- & +6.4\% (33.5) & +7.2\% (37.7) \\
Both on 4 disks & 34\,GiB & --- & +2.6\% (32.0) & +2.6\% (36.1) \\
Both on 1 disk, cache on layer~14 only & 34\,GiB & 0.5M & +5.6\% (33.3) & +7.1\% (37.7) \\
Both on 1 disk, cache split evenly & 34\,GiB & 1M & \textbf{+1.2\%} (31.8) & +4.8\% (37.0) \\
Both on 4 disks, cache split evenly & 34\,GiB & 1M & \textbf{+0.7\%} (31.2) & \textbf{+1.7\%} (35.8) \\
\bottomrule
\end{tabular}
\end{table*}

Moving only layer~14 to NVMe adds no measurable overhead, even without a cache.
Its 2--4\,ms of reads finish well within the window, and p99 step time stays unchanged.
Almost all overhead comes from layer~1.
With both tables on one disk, caching layer~14 with 0.5M rows reduces overhead only from 6.4\% to 5.6\%.
The full 1M rows reduce it to 4.9\% (Figure~\ref{fig:alloc}).
Splitting the same 1M rows between the two layers removes four fifths of it (1.2--1.3\%).
Chat's misses are mostly first-seen rows, so four disks, one per rank with four I/O threads each, help it more than the cache split.
Together they bring chat to +1.7\% and agents to +0.7\%.

\subsection{Allocating the cache}
\begin{figure}[t]\centering
\includegraphics[width=\linewidth]{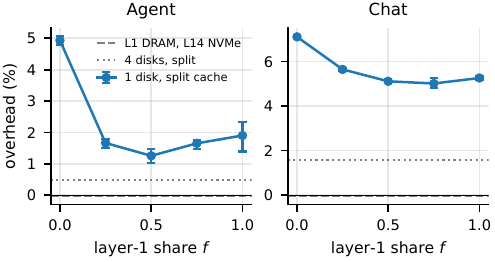}
\caption{Decode overhead vs.\ the share of the cache given to layer~1 ($B{=}128$, three runs, bars show min/max). Dashed: layer~1 in DRAM, layer~14 on NVMe with no cache. Dotted: four disks with four I/O threads and an even split.}
\label{fig:alloc}\end{figure}

Figure~\ref{fig:alloc} varies the share $f$ of a 1M-row-per-rank cache given to layer~1, with three runs per point in shuffled order.
For agents, every $f$ from 0.25 to 1 yields overhead between +1.3\% and +1.9\%. Giving everything to layer~14 ($f{=}0$) costs +4.9\%.
Chat shows the same shape at a higher overhead (+5.0--5.6\% vs.\ +7.1\%) because most of its misses are first-seen rows.
Layer-14-only caching does poorly because it leaves the short-window layer exposed, although the two layers have similar hit rates.
Giving layer~1 between a quarter and all of the cache yields similar performance, so layer-1-only caching and an even split are both effective.
An online hill-climbing controller moves away from $f{=}0$, and from $f{=}0.5$ it ends within 0.15\,ms of the fixed split, so we keep $f{=}0.5$.

Allocation affects overhead more than size does.
With a quarter of the budget (0.25M rows per rank, 33\,MB per layer), an even split adds 1.8\% for agents.
The same cache on layer~14 alone adds 4.8\%.
Doubling the budget brings the split to 0.4\%; doubling again has no effect.

\subsection{Measured prefetch windows}
We measure windows directly with a small probe that works on any SGLang model without changing model code.
It hooks the decoder layers and queues one host node per layer on a side stream at each step's start.
Layer $k$ waits on its node, which completes $X$ after the step starts.
The probe runs inside the CUDA graph, so we can change $k$ and $X$ at runtime and see where the step starts to slow down.

On V4.1, delaying layer~14's lookup by 3, 6 or 9\,ms changes the step by at most 0.18\,ms.
Delays of 12 and 15\,ms add 2.65 and 5.61\,ms, putting $W_{14}$ at about 9.3\,ms.
The model's estimate of 13 layers' worth of compute is 9.5\,ms.
For layer~1, delays of 0.2 and 0.4\,ms cost about 0.1\,ms.
From 0.6\,ms on, every extra millisecond shows up almost one for one, so $W_1$ is only about 0.3\,ms, roughly 30$\times$ shorter than $W_{14}$.

We then delay both layers at once, with six pairs not used to set any parameter, and predict the slowdown from Eq.~\ref{eq:exposure} using only the two measured windows.
The prediction is within 0.12\,ms RMSE (Table~\ref{tab:both}).
Adding the two layers' excesses misses by 1.29\,ms because a layer~1 stall gives layer~14 extra time.
For example, raising the layer-1 delay from 1.5 to 2.5\,ms with layer~14 at 15\,ms barely changes the step (+5.46 vs.\ +5.54\,ms).

\begin{figure}[t]\centering
\includegraphics[width=\linewidth]{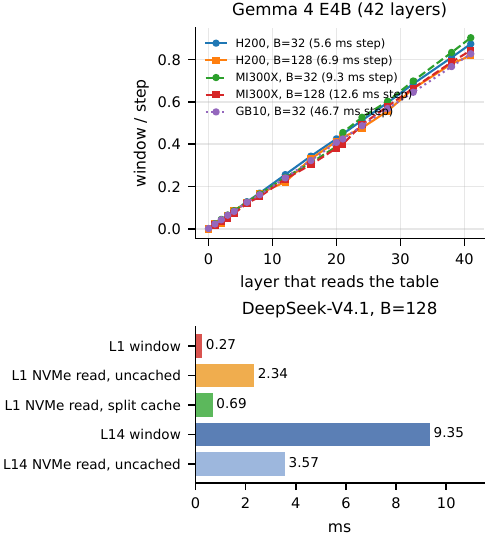}
\caption{Measured prefetch windows. Top: Gemma~4 E4B under SGLang, 16 of 42 layers, on H200, MI300X and GB10, as a fraction of each platform's decode step. Bottom: V4.1's two Engram layers next to measured mean NVMe read times per step. A window is estimated as $X-\Delta$ at the first injected delay $X$ whose step-time increase $\Delta$ exceeds the noise threshold.}
\label{fig:windows}\end{figure}

\begin{table}[t]
\centering\footnotesize
\caption{Held-out check: both layers delayed at once ($B{=}128$, agent, mean of two runs). All times in ms. Predictions use only the measured windows $W_1{=}0.27$\,ms and $W_{14}{=}9.35$\,ms.}
\label{tab:both}
\resizebox{\columnwidth}{!}{%
\begin{tabular}{rrrrr}
\toprule
L1 delay & L14 delay & Measured & Eq.~\ref{eq:exposure} & Sum of excesses \\
\midrule
0.4 & 9 & +0.12 & 0.13 & 0.13 \\
0.9 & 9 & +0.64 & 0.63 & 0.63 \\
0.9 & 12 & +2.52 & 2.65 & 3.28 \\
1.5 & 12 & +2.51 & 2.65 & 3.88 \\
1.5 & 15 & +5.46 & 5.65 & 6.88 \\
2.5 & 15 & +5.54 & 5.65 & 7.88 \\
\bottomrule
\end{tabular}}
\end{table}
On Gemma~4 E4B (42 layers), the window grows linearly with depth, and layer~0 has none (Figure~\ref{fig:windows}).
The slope tracks per-layer compute: about 0.12--0.14\,ms per layer on an H200 and 0.20--0.26\,ms on an MI300X.
On a DGX Spark (GB10, unified memory), where a step takes 47\,ms at $B{=}32$, it is 0.94\,ms.
Normalized by step time, the curves from the three platforms nearly coincide: layer~$k$'s window is about $0.85$--$0.92\,k/L$ of a step (16 of the 42 layers probed).

\subsection{Ablations}
Table~\ref{tab:abl} turns off one design choice at a time (one disk, even split, two runs each; the layer-14-only baseline has three).
On chat, which misses more, interleaving the two layers' reads instead costs 0.8\,ms per step. With the cache on layer~14 only, it costs another 0.5--0.6\,ms.
Deduplicating pages and rows within a step saves 0.2--0.4\,ms.

\begin{table}[t]
\centering\small
\caption{Ablations: mean decode step (ms) at $B{=}128$, one disk.}
\label{tab:abl}
\begin{tabular}{lrr}
\toprule
 & Agent & Chat \\
\midrule
Full design (even split) & 29.62 & 35.30 \\
\quad $-$ layer-1-first I/O & 29.86 & 36.09 \\
\quad $-$ in-step dedup & 29.86 & 35.66 \\
\midrule
L14-only cache & 30.85 & 36.09 \\
\quad $-$ layer-1-first I/O & 31.34 & 36.69 \\
\bottomrule
\end{tabular}
\end{table}
Without CUDA graphs each step is about 4$\times$ longer (127--128\,ms) and every NVMe configuration falls within $\pm$1\% of DRAM; capturing the I/O path in the graph keeps the launch overhead out.

\paragraph{Page-cache baseline.}
SGLang's file backend maps the table and lets the OS page cache hold hot rows; it needs GPU access to pageable memory, which our H200s lack, so we emulate it with the same I/O path without O\_DIRECT, WILLNEED hints and no GPU cache, in a cgroup that caps the page cache (four disks, three runs).
With the backend's default 8\,GiB the step is 4.2\% (agent) and 10.8\% (chat) slower than DRAM; uncapped, the page cache grows to 90\,GiB and is still 1.4\% and 2.6\% slower, against 0.1\% and 1.6\% for SlackTier with a 1\,GB GPU cache.

\subsection{Sensitivity to load}
\begin{figure}[t]\centering
\includegraphics[width=\linewidth]{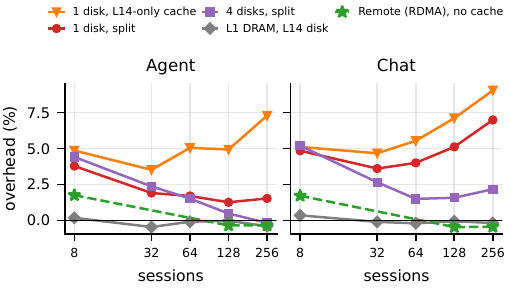}
\caption{Decode step overhead vs.\ concurrent sessions, relative to the DRAM table at the same load. NVMe: one run per point (three at 128); RDMA: measured at 8, 128 and 256 sessions in a separate campaign against its own DRAM baseline.}
\label{fig:conc}\end{figure}

Figure~\ref{fig:conc} varies the number of concurrent sessions from 8 to 256.
Keeping layer~1 in DRAM and moving layer~14 to disk stays within $\pm$0.5\% at every load.
With both tables on one disk, the gap between the two cache allocations grows with load.
At 256 sessions, a layer-14-only cache costs +7.3\% for agents and +9.0\% for chat, while an even split costs +1.5\% and +7.0\%.
Four disks with an even split cost $-$0.2\% and +2.2\%.
At 8 sessions, every configuration that reads layer~1 from disk adds about 0.4--0.6\,ms per step.
Each step reads only about 190 rows, so layer~1 waits for one disk round trip that neither caching nor more disks can hide.
Remote memory lowers this to 1.7--1.8\%, with no measurable overhead from 128 sessions up.

\paragraph{Prefill and TTFT.}
Prefill reads use the same path, which deduplicates within each 8K-token chunk (agent chunks keep about a third of their rows).
Measured TTFT shows little exposed read time.
We sent the same pre-tokenized prompts of 4K, 16K and 32K tokens, alone and eight at a time, to unmodified SGLang and to SlackTier (two blocks, three repetitions per point).
SlackTier ran with the tables in DRAM, with layer~14 on disk, and with both tables on four disks.
In all three, median TTFT is within 0.04\,s of unmodified SGLang at every point (mean over two blocks).
At 32K tokens it is 1.80 vs.\ 1.81\,s alone, and 8.63--8.64 vs.\ 8.67\,s eight at a time.

\paragraph{Idle gaps between turns.}
Real agents pause between turns while tools run, and other sessions can evict a paused session's rows.
Adding a 1--6\,s pause after every turn leaves the layer-1 hit rate at 84--86\%.
We replay a fixed sequence with these pauses at 256 sessions (two blocks).
Clients see 151.9\,ms per output token from unmodified SGLang and 150.1\,ms from SlackTier with the tables in DRAM.
With both tables on four disks, they see 149.5\,ms.
Median TTFT is 5.6, 5.3 and 5.3\,s.

\paragraph{Reusing the freed memory.}
The freed memory can hold KV cache.
We shrink the GPU KV pool to 1.5M tokens so that sessions paused for tools get evicted.
We run 256 agent sessions against four servers, five runs each (Table~\ref{tab:hicache}).
Moving the tables to disk leaves every client metric unchanged.
Giving the freed memory to SGLang's hierarchical cache raises the prefix hit rate by 14 points.
It cuts median p90 TTFT from 38 to 24\,s and raises median throughput by 36\% (20--44\% across runs, against 701--836 tokens/s without it).
Reaching the same result with the tables in DRAM takes 408\,GiB.
The gain depends on the GPU KV budget: with a 3M-token pool, 77\% of prompt tokens already hit without a host KV cache, and adding one changes neither the hit rate nor throughput. Moving the tables to disk again leaves every client metric unchanged.

\begin{table}[t]
\centering\footnotesize\setlength{\tabcolsep}{2.5pt}
\caption{Spending the freed memory on host KV cache: 256 agent sessions with tool pauses, GPU KV pool fixed at 1.5M tokens. RSS: resident set size. Medians of five runs.}
\label{tab:hicache}
\resizebox{\columnwidth}{!}{%
\begin{tabular}{llrrrrr}
\toprule
Tables & Host KV & RSS & Prefix hit & TTFT p50 & TTFT p90 & Tokens/s \\
\midrule
DRAM & --- & 223\,GiB & 63\% & 13.6\,s & 38.1\,s & 702 \\
4 disks & --- & 34\,GiB & 62\% & 14.5\,s & 37.6\,s & 700 \\
4 disks & 185\,GiB & 220\,GiB & 76\% & 10.1\,s & 24.1\,s & 956 \\
DRAM & 185\,GiB & 408\,GiB & 75\% & 11.2\,s & 25.5\,s & 954 \\
\bottomrule
\end{tabular}}
\end{table}

\subsection{Client-side comparison with unmodified SGLang}\label{sec:client}
We replay a fixed request sequence against unmodified SGLang and SlackTier.
Every configuration sees the same sessions, prompts and output lengths (up to eight turns per session, $B{=}128$).
Turns generate 64 tokens.
Each block runs our configurations once in random order.
Unmodified SGLang ran in three separate blocks on the same sequence.
Table~\ref{tab:client} reports the mean over blocks and the range.
Prefill uses the stock GPU gather when both tables are in DRAM and our tier otherwise.

\begin{table}[t]
\centering\footnotesize\setlength{\tabcolsep}{2.5pt}
\caption{Client-side cost relative to unmodified SGLang with the tables in host DRAM. Fixed request sequence, $B{=}128$. Mean time per output token (TPOT) over 4 blocks [min, max], and time to finish the whole sequence. Unmodified SGLang: 81.6\,ms (agent) and 44.2\,ms (chat), 3 blocks.}
\label{tab:client}
\resizebox{\columnwidth}{!}{%
\begin{tabular}{lrrrr}
\toprule
 & \multicolumn{2}{c}{Agent} & \multicolumn{2}{c}{Chat} \\
SlackTier, tables in & TPOT & Total & TPOT & Total \\
\midrule
Host DRAM & +1.0\% [$-$1.0, 3.0] & +0.1\% & +0.5\% [$-$0.4, 0.7] & $-$0.2\% \\
L1 DRAM, L14 NVMe & $-$0.8\% [$-$2.8, 0.3] & +0.1\% & +2.7\% [0.6, 8.1] & +3.1\% \\
Both on 1 disk, split & +3.6\% [2.5, 5.9] & +4.3\% & +6.7\% [6.1, 7.4] & +7.2\% \\
Both on 4 disks, split & $-$0.5\% [$-$2.2, 0.5] & +0.0\% & +1.4\% [1.0, 1.9] & +0.6\% \\
Both remote (RDMA) & +0.2\% [$-$1.2, 1.0] & $-$0.4\% & $-$0.1\% [$-$0.4, 0.1] & $-$0.8\% \\
\bottomrule
\end{tabular}}
\end{table}

With four disks or remote memory, mean time per token is at most 1.4\% above unmodified SGLang.
Total time is at most 0.6\% above it.
The p95 time per token is within 2.4\% (agent 154.0 vs.\ 155.4--157.6\,ms; chat 72.8 vs.\ 72.8--73.9\,ms).
Every block completes the same 1016 (agent) or 1338 (chat) requests.
The serving process uses 34 instead of 212\,GiB of host memory.
On one disk, prefill also reads from disk, and chat, which prefills more per output token, is 7\% slower.
Median TTFT stays within 0.2\,s of unmodified SGLang in every configuration.

\subsection{A second model: Qwen3.8 PLE}
Qwen3.8-Flash-Next (48 layers, 180B parameters in BF16) has one hashed n-gram table, its per-layer embedding (PLE).
The table has 320M rows of 320\,B and is read at layer~1.
Our probe gives that layer a window of 0.27\,ms, the same as V4.1's layer~1.
Windows grow roughly linearly to 14.5\,ms at the last layer.
SGLang pins the table in host memory by default, using 201\,GiB for the serving process.
With the table in HBM, the process uses 74\,GiB.

We hooked SlackTier into Qwen's PLE gather with no other changes.
An in-server check ran the original gather next to ours and found no mismatching row over 1.1M checked lookups, including cache hits and evictions.
Table~\ref{tab:qwen} shows per-token decode time at 128 sessions (mean of two runs).
Qwen's tier runs at its own prefetch point, one layer ahead.
As with V4.1's layer~1, disks alone help little, and a cache brings decode time close to the host-memory baseline.
With four disks and a 1M-row cache, decode is 3.4\% slower for agents and 2.4\% for chat.
The cache serves 93\% and 57\% of lookups.

\begin{table}[t]
\centering\footnotesize
\caption{Qwen3.8-Flash-Next, 128 sessions: per-token decode time (mean of two runs' medians) relative to the table in host memory (36.8\,ms agent, 24.8\,ms chat).}
\label{tab:qwen}
\begin{tabular}{lrr}
\toprule
PLE table location & Agent & Chat \\
\midrule
1 disk, no cache & +9.4\% & +7.9\% \\
4 disks, 4 I/O threads, no cache & +5.8\% & +3.2\% \\
1 disk, 1M-row cache & +4.5\% & +4.8\% \\
4 disks, 4 I/O threads, cache & +3.4\% & +2.4\% \\
\bottomrule
\end{tabular}
\end{table}

\subsection{Sharing and remote tiers}
To emulate $k$ instances sharing a disk, each page read issues $k{-}1$ extra reads of other pages on the same disk.
With four instances sharing one disk, the uncached configuration degrades by 23\% (agent) and 26\% (chat).
The split cache keeps agent at +5.5\%, and four disks with per-disk threads keep it at +2.7\%.
\paragraph{Tables in another node's memory.}
SlackTier can also read both tables from the memory of a second H200 node.
A small server lays out the tables (195\,GiB) in the same 4\,KiB pages as on disk and registers them with one 400\,Gb/s InfiniBand NIC per rank.
The only change in the tier is that each disk read becomes a one-sided RDMA read, issued from four threads over two queue pairs each.
The device-side check found no mismatching row over 3.9M decode and 18.8M prefill lookups per rank and layer.
An uncached step at $B{=}128$ reads about 600--740 pages per rank and layer in 0.21--0.25\,ms, ten times faster than NVMe.
Layer~14's reads are fully hidden.
Layer~1's host-side wait is about 0.2\,ms, but it overlaps layer~0's kernels, which are already queued.
An Nsight Systems trace shows only 4\,$\mu$s of GPU idle time before each Engram gather, the same as with the tables in DRAM.
At $B{=}128$ and 256, every remote placement, cached or not, is within $-0.7$\% to $+0.0$\% of the host-DRAM table, with unchanged p99 step time.
With no table loaded locally, the serving process uses 34\,GiB of host memory instead of 223\,GiB.
Only at $B{=}8$, where a step takes 11\,ms, does layer~1's wait show.
Uncached overhead is +1.8\% (agent) and +1.7\% (chat), versus +1.4\% and +1.6\% with a 1M-row cache.
Two serving instances, each on four GPUs of another node, can read one copy of the tables at once.
Their time per output token is within 2.2\% of a single instance with the tables in its own DRAM (two blocks each, fixed request sequence).
Uncached remote reads take as long as with one instance (0.16--0.22\,ms per layer and step).
An earlier emulation that delays completion by 15\,$\mu$s at 100\,Gb/s or 120\,$\mu$s at 25\,Gb/s agrees for layer~14 ($\le$0.3\%).

\paragraph{Lossy skipping.} An optional mode that skips rows never seen in about 75M tokens of traces lowers agent overhead from +1.6\% to +0.5\%. Quality results are in Appendix~\ref{app:lossy}.

\section{Discussion}\label{sec:disc}

\paragraph{Why not other weights.}
The same method does not apply to dense or expert weights.
A decode step reads nearly all of V4.1's non-Engram weights, about 307\,GB in the released checkpoint.
Even at the limit of four PCIe~5 x16 links, about 64\,GB/s each, reading them takes about 1.2\,s per step, against measured decode steps of 11--43\,ms.
An Engram step at $B{=}128$ reads about 12\,MB of 4\,KiB pages per layer, and which pages is known in advance.
Systems that stream weights from storage~\cite{sheng2023flexgen,alizadeh2024flash} therefore target offline batches or single-user devices.

\paragraph{Unified memory.}
On unified-memory devices a host-resident table takes memory from weights and KV cache. Qwen3.8's NVFP4 release (124\,GiB, including a 48\,GiB PLE table) runs out of memory while loading on a DGX Spark (GB10, 128\,GB) with the table in host memory, under SGLang and vLLM.
With the table memory-mapped from local NVMe (a community vLLM patch), it loads in 74\,GiB and keeps 17\,GiB (0.58M tokens) for KV cache, decoding at 51--96\,ms per token from 1 to 8 requests.
That patch reads rows from the page cache only when layer~1 needs them, so 3--7\,ms of host time per lookup, 4--6\% of each step, sits on the critical path.
GB10 windows are also wider: 0.94\,ms per layer for Gemma~4, seven times an H200's.

\paragraph{Who needs the memory.}
Our H200 node has 1.8\,TiB of DRAM, and freed memory pays off as host KV cache only under a tight GPU KV budget (\S\ref{sec:eval}).
Nodes with less DRAM per GPU, several instances per node, and unified-memory devices lose KV cache or fail to load with the tables in memory~\cite{aetherixengram,vllm62238}, and DRAM costs one to two orders of magnitude more per byte than NVMe.

\paragraph{Composing with quantization.}
Quantized tables~\cite{nvidianvfp4,han2026twobits} keep the access pattern, so the rule is unchanged, while a 4-bit table roughly halves its footprint.
Appendix~\ref{app:disc} discusses larger tables, on-demand fetching and disaggregated serving.

\paragraph{Scope.}
End-to-end runs use two models on 4$\times$H200 nodes; windows on other hardware come from the probe, and shared disks are emulated.
New tables, tiers and hardware need each layer's window and read time under the target load, plus the tables' miss curves.

\section{Related Work}

\paragraph{Serving lookup parameters.}
Engram~\cite{cheng2026engram} observes that deterministic addressing allows runtime prefetching from host memory.
SGLang and vLLM keep both tables in pinned host DRAM, which frees HBM for KV cache~\cite{lmsysv41}.
\citet{ma2026cxlengram} define the zero-stall condition (lookups finish within the compute of the preceding layers), note that early layers are tight, and find on a proxy model that a CXL pool meets it while an RDMA pool does not.
Table~\ref{tab:related} lists the closest work.
Low-bit tables~\cite{nvidianvfp4,chung2026lowbit} halve their memory, and DwarfStar and DeepInfer-Mac read rows from SSD on single laptops or workstations~\cite{dwarfstar,li2026deepinfermac}.
TF-Engram~\cite{tfengram2026} and STAPLE~\cite{staple2026} put other memory tables in host or SSD tiers with predictive prefetch or a frequency cache.
Since V4.1's release, deployments have served both tables from SSD~\cite{semianalysisengram,zheng2026nowait}, from NVMe or DRAM chosen per device~\cite{aetherixengram}, from a remote host over RDMA over Converged Ethernet (RoCE)~\cite{plirdma}, or from peer HBM~\cite{vllm62091}.
Each puts both tables on one tier; we choose the tier per layer and split the cache by window. Concurrently, a blog post models both deadlines for single-stream decode from SSD latencies~\cite{petersson2026engramssd}; we measure them under batched serving and evaluate end to end.

\begin{table}[t]
\centering\footnotesize\setlength{\tabcolsep}{2pt}
\caption{Closest work on serving lookup tables.}
\label{tab:related}
\begin{tabular}{p{0.23\linewidth}p{0.33\linewidth}p{0.37\linewidth}}
\toprule
Work & Shows & This paper adds \\
\midrule
\citet{ma2026cxlengram} & Zero-stall condition; CXL pool near DRAM on a proxy model & Per-layer placement and cache split on released V4.1; NVMe and RDMA served end to end \\
MiaAI PR~\cite{miaaipr7} & NVMe row cache for both layers on GB10; layer-1 stalls larger than layer-14 & Measured windows that explain it; tier and cache chosen per layer by window \\
vLLM \#59958~\cite{vllm59958} & Qwen PLE read from disk, compatible with CUDA graphs & Session locality; disk contention between layers \\
vLLM \#57023~\cite{vllm57023} & V4.1 tables in a Mooncake store, overlapped lookups & Placement by window; shared remote table measured with two instances \\
\citet{han2026twobits} & 2-bit tables; static hot-row cache & Static vs.\ session caches on interleaved agent traffic \\
\bottomrule
\end{tabular}
\end{table}

\paragraph{Tiered placement of model state.}
FlexGen~\cite{sheng2023flexgen} places weights, KV cache and activations across GPU, CPU and disk with a cost model that maximizes offline throughput. DeepSpeed-Inference~\cite{aminabadi2022deepspeed} offloads weights for large-model inference, and ZeRO-Infinity~\cite{rajbhandari2021zeroinf} offloads training state to CPU and NVMe.
Select-N~\cite{ma2025selectn} uses deterministic per-layer compute time to decide how much state to offload under a latency service-level objective (SLO), and ITME~\cite{jang2026itme} prefetches weights and KV from a CXL tier.
These systems read each layer's tensors in full and optimize bandwidth; our tables need a few sparse \emph{rows} per step, each layer's by its own deadline.

\paragraph{Predictive offloading.}
LLM in a Flash~\cite{alizadeh2024flash} and PowerInfer~\cite{song2024powerinfer} offload feed-forward (FFN) neurons and predict which will be active. Pre-gated MoE~\cite{hwang2024pregated} and MoE-Infinity~\cite{xue2024moeinfinity} offload experts and fetch them ahead of use, by early gating or by caching experts according to activation history.
Those indices depend on activations and must be predicted or, as in Pre-gated MoE, made known one layer early; Engram's come from the input.

\paragraph{Embedding caches.}
Recommendation models~\cite{naumov2019dlrm} cache hot embedding rows on GPU~\cite{wang2022hugectr,song2023ugache}, process them on SSDs~\cite{wilkening2021recssd} or prefetch them across batches~\cite{agarwal2023bagpipe}, and BaM~\cite{qureshi2023bam} lets GPUs read storage directly.
Their embeddings are consumed at the input, with no per-layer deadlines; we find \emph{session} locality that static placement misses.

\paragraph{KV-cache tiering.}
vLLM~\cite{kwon2023vllm}, Mooncake~\cite{qin2025mooncake}, LMCache~\cite{liu2025lmcache}, CacheGen~\cite{liu2024cachegen} and HiCache page or move KV cache in prefix blocks; Engram reads are per-step rows, but long-window layers could use the same remote pools, as our RDMA tier does.

\bibliography{refs}
\bibliographystyle{mlsys2025}

\section*{AI Usage Disclosure}
We used AI assistants (Anthropic Claude and OpenAI Codex) in this work.
The authors proposed the research question and the overall idea, and led the experiments: they designed the experimental plan, chose the models, workloads and hardware, decided which follow-up experiments to run, and judged which results support which claims.
AI assistants helped check the topic against prior work by searching for and summarizing related papers, pull requests and model releases.
Under the authors' direction, they helped refine the experimental plan and assisted with code, experiment runs, figures and tables, and writing.
The authors reviewed all AI-assisted code and text, checked the reported numbers and citations, and take full responsibility for the content.

\appendix
\section{Full Results}
\label{app:full}
Table~\ref{tab:full} (next page) lists every configuration of the cache-allocation sweep behind Figure~\ref{fig:alloc} and Table~\ref{tab:main}. All runs use $B{=}128$ and three shuffled runs per configuration on one server process. Layer-1 hit rates rise from 0.81 to 0.87 as $f$ grows from 0.25 to 1 for agents, while chat stays at 0.33--0.43; from $f{=}0.25$ to $f{=}0.75$ the mean step changes by at most 0.22\,ms, so an even split is a safe default.

\begin{table*}[!t]
\centering\small
\caption{All configurations of the allocation sweep (an earlier campaign than Table~\ref{tab:main}, which is why the baselines differ slightly), $B{=}128$, three shuffled runs on one server process. Step: mean decode step in ms (max$-$min across runs). p99: 99th-percentile step in ms. L1 hit: fraction of layer-1 lookups served by the GPU cache. Cache budget 1M rows per rank; $f$ is the share given to layer~1.}
\label{tab:full}
\begin{tabular}{lrrrrrr}
\toprule
 & \multicolumn{3}{c}{Agent (OpenHands)} & \multicolumn{3}{c}{Chat (WildChat)} \\
Configuration & Step & p99 & L1 hit & Step & p99 & L1 hit \\
\midrule
Host-DRAM table & 29.40 (0.02) & 31.1 & --- & 33.69 (0.08) & 35.1 & --- \\
L1 DRAM, L14 1 disk, no cache & 29.39 (0.00) & 31.1 & --- & 33.67 (0.11) & 35.1 & --- \\
1 disk, $f{=}0$ & 30.85 (0.08) & 33.0 & 0.00 & 36.09 (0.02) & 37.7 & 0.00 \\
1 disk, $f{=}0.25$ & 29.89 (0.09) & 31.9 & 0.81 & 35.60 (0.04) & 37.2 & 0.33 \\
1 disk, $f{=}0.5$ & 29.77 (0.13) & 31.6 & 0.85 & 35.42 (0.05) & 37.1 & 0.39 \\
1 disk, $f{=}0.75$ & 29.89 (0.09) & 31.7 & 0.86 & 35.38 (0.15) & 37.0 & 0.41 \\
1 disk, $f{=}1$ & 29.96 (0.28) & 31.8 & 0.87 & 35.47 (0.05) & 37.1 & 0.43 \\
4 disks, 4 threads, no cache & 30.09 (0.06) & 32.0 & --- & 34.62 (0.14) & 36.0 & --- \\
4 disks, 4 threads, $f{=}0.5$ & 29.54 (0.13) & 31.3 & 0.85 & 34.22 (0.04) & 35.7 & 0.39 \\
1 disk, controller from $f{=}0.5$ & 29.92 (0.05) & 31.9 & 0.85 & 35.32 (0.10) & 37.0 & 0.42 \\
1 disk, controller from $f{=}0$ & 30.06 (0.18) & 31.9 & 0.69 & 35.60 (0.05) & 37.2 & 0.31 \\
\bottomrule
\end{tabular}
\end{table*}

\section{Experimental Details}
\label{app:setup}
\paragraph{Hardware.} One cloud node with 8$\times$H200 (we use 4), 1.8\,TiB host memory, eight local 3.5\,TB NVMe SSDs exposed as raw block devices (we use up to four). Gemma~4 probes also ran on one AMD Instinct MI300X (gfx942) (8-GPU node).
\paragraph{Software.} SGLang 0.5.22 (pinned container image), DeepSeek-V4.1-Flash-1017 checkpoint, TP4/EP4, FP8 KV cache, decode CUDA graphs, decoder SWA bounded replay, chunked prefill 8192. Our changes are a patch to the Engram embedding's gather and two hooks in the model (after hashing and after weight loading).
\paragraph{Table staging.} Each rank's quarter of each table is written to its own region of the disk(s) as 264\,B records, 15 per 4\,KiB page. We verify 200 random rows per (layer, rank, disk) against the checkpoint before every run, and discard and verify the regions afterward.
\paragraph{Correctness.} An in-server checker runs the original gather next to ours on every lookup and counts mismatching rows on real tokens. Across DRAM, 1-disk, 4-disk, cached and tiny-cache (forcing evictions) configurations, it found 0 mismatches over $\sim$790K lookups in prefill and decode.
\paragraph{Workloads.} Agent: 512 OpenHands trajectories from SWE-rebench, from repositories never used to build any statistic, average 36 turns. Chat: 512 WildChat conversations from a held-out shard. Each assistant turn is regenerated with 128 new tokens (ignore EOS) from the original history, so prefixes hit the radix cache across turns.
\paragraph{Window probe.} A forward pre-hook on the first decoder layer queues one host-function node per layer on a side stream; layer $k$'s pre-hook makes the main stream wait on node $k$, which busy-waits until $X$ after the step began if $k$ is the target and returns immediately otherwise. The configuration is read from a file at runtime, so one captured graph serves every $(k, X)$. We fit the window as $X - \Delta$ at the first $X$ where the step grows by more than a noise threshold: 0.3\,ms on H200 and MI300X, and on GB10, whose steps are noisier, the larger of 0.3\,ms, three standard deviations of the baseline step, and 1\% of the step (0.51\,ms).

\section{Access Characterization Details}
\label{app:char}
\begin{table}[h]
\centering\footnotesize\setlength{\tabcolsep}{3pt}
\caption{LRU hit rate (\%, all lookups) vs.\ cache capacity, as a fraction of both tables (768M rows), 256 interleaved sessions. Last column: decode lookups only at 5\%.}
\label{tab:hitfull}
\resizebox{\columnwidth}{!}{%
\begin{tabular}{lrrrr}
\toprule
Workload & 0.2\% & 1\% & 5\% & Decode @5\% \\
\midrule
OpenHands agent, cold & 62 & 78 & 86 & 86 \\
OpenHands, warmed on other repos & 62 & 78 & 86 & 86 \\
SWE-agent (one trajectory per issue) & 59 & 85 & 88 & 89 \\
WildChat chat, cold & 30 & 45 & 53 & 47 \\
\bottomrule
\end{tabular}}
\end{table}

\paragraph{Static hot sets.} Ranking rows by frequency in a fit split and pinning the top 5\% covers 29--39\% of chat and 53--65\% of SWE-agent lookups as the fit data grows from 3M to 60M tokens per domain. A public hot list built from 1.1B tokens of assistant traffic~\cite{botlabhot} reaches 74\% at 20M rows on its own held-out data, but only 21--42\% on our workloads.

\paragraph{By n-gram order.} Static coverage falls with the order of the n-gram: 83--91\% for 2-grams, 48--70\% for 3-grams, 28--54\% for 4-grams. Rows never seen in about 75M tokens of fit data make up 7--11\% of agent lookups and 14--23\% of chat lookups.

\paragraph{Batch reuse.} Within an 8K-token window of the trace, 33\% (SWE-agent) to 71\% (chat) of row lookups are distinct; among 256 sampled decode tokens, 95--99.6\% are.

\paragraph{Hash heads are not redundant.} Reading the 11,520 sampled layer-14 rows and the projection weights, the eight heads of the same n-gram order have nearly orthogonal value contributions (cosine $\approx0$); dropping any one head changes that n-gram order's projected value contribution by about 35\% in norm.

\paragraph{Speculative prefetch.} Using prompt-lookup drafting as a proxy for a draft model (one or four candidates), prefetching the predicted next token's rows converts only 1.7--7.2\% of lookups from exposed misses to prefetched hits, because the drafter misses mostly on the same novel tokens that miss the cache. It fetches 1.1--7.9 extra n-gram keys per token (eight rows per key and layer).

\paragraph{Offline packing.} Storing each frequent 4-gram context's 24 rows contiguously (two pages instead of up to 24) reduces page reads behind a 2M-row-per-layer cache by 22\% for agents and 6--8\% for chat, saturating at 50\,GB (agent) to 200\,GB (chat) of extra disk; under 10\% of missing tokens hit a packed context.

\paragraph{Second model.} Reproducing Qwen3.8-Flash-Next's PLE hash (2/3-grams, 8 heads, 320M rows), session LRU reaches 86--87\% on agent traces and 54--60\% on chat, and a prefill-to-decode handoff of the session's rows covers 76\% of agent decode lookups, close to V4.1's 73--76\%.

\section{Lossy skipping}\label{app:lossy}
Optionally, rows absent from a seen-set built on about 75M tokens can be skipped on a miss, reducing agent overhead from +1.6\% to +0.5\%.
Teacher-forced negative log-likelihood (NLL) on agent text changes by $-0.007$ to $-0.009$ (chat $+0.010$), while dropping a random matched fraction costs $+0.021$.
HumanEval and MBPP change by $-1.8$ to $-4.3$ points, within the $\pm6$-point confidence interval of our 164/257-problem runs; GSM8K accuracy on the 149 problems where both configurations produce a final answer is 93.3\% vs.\ 92.6\%.
We keep skipping as an opt-in mode for agent workloads.

\section{Further discussion}\label{app:disc}
\paragraph{When tables grow.}
The Engram paper reports that quality keeps improving as tables grow. At a terabyte, a host-DRAM table is no longer an option on most nodes, while NVMe capacity is cheap.
Deep-layer tables can live on disk or a remote tier when their reads fit the window; early-layer tables need a fast tier or a cache sized for their sessions' working set. Growing a hashed table does not change the number of lookups per token (24 per layer); if, as in our traces, misses are mostly first-seen n-grams, the number of missing rows per step changes little as well. Under Eq.~\ref{eq:exposure}, the NVMe configuration's overhead therefore stays roughly constant, and its host footprint grows only with the CPU directory (4\,B per row). On our node, a 2$\times$ table would need about 414\,GiB of host memory in DRAM mode and 224\,GiB with layer~14 on NVMe, while the full-NVMe process grows only by its directory (4\,B per row, about 3\,GiB per additional 768M rows) from today's 34\,GiB; at 5$\times$, DRAM mode needs about 983\,GiB, more than half the node, while full NVMe needs about 46\,GiB plus 1\,TB of disk.

\paragraph{Disaggregated serving.}
In prefill/decode disaggregation~\cite{zhong2024distserve,qin2025mooncake}, we find that a session's Engram rows from prefill predict its decode lookups well.
We simulate a decode instance that receives these rows with the KV cache and then caches its own rows.
It serves 73--76\% of agent decode lookups from cache; the prompt's rows are a few MiB per layer.

\paragraph{Fetching on demand.}
The simplest offload, fetching a layer's rows synchronously when it needs them, puts the transfer on the critical path. In a quick test with Hugging Face's eager runtime, moving Gemma~4's per-layer embedding table to CPU memory this way, a recipe the community uses to save VRAM, slowed decode by 5.6$\times$ at 32 concurrent requests. Issuing lookups early and caching the short-window table avoids this.

\end{document}